\documentclass[aps,prl,twocolumn,groupedaddress,superscriptaddress,amssymb,amsmath,showpacs]{revtex4}
\usepackage{dcolumn}
\usepackage{graphicx}

\begin{document}


\title{Theory of Magnetic Edge States in Chiral Graphene Nanoribbons}

\author{Oleg V. Yazyev}
\affiliation{Department of Physics, University of California, Berkeley, California 94720, USA}
\affiliation{Materials Sciences Division, Lawrence Berkeley National Laboratory, Berkeley, California 94720, USA}
\author{Rodrigo B. Capaz}
\affiliation{Department of Physics, University of California, Berkeley, California 94720, USA}
\affiliation{Instituto de F\'isica, Universidade Federal do Rio de Janeiro, Caixa Postal 68528, Rio de Janeiro, RJ 21941-972, Brazil}
\author{Steven G. Louie}
\affiliation{Department of Physics, University of California, Berkeley, California 94720, USA}
\affiliation{Materials Sciences Division, Lawrence Berkeley National Laboratory, Berkeley, California 94720, USA}

\date{\today}


\begin{abstract}
Using a model Hamiltonian approach including electron-electron interactions, we systematically investigate
the electronic structure and magnetic properties of chiral graphene nanoribbons. We show that the presence
of magnetic edge states is an intrinsic feature of smooth graphene nanoribbons with chiral edges, and 
discover a number of structure-property relations. Specifically, we study the dependence of magnetic moments
and edge-state energy splittings on the nanoribbon width and chiral angle as well as the role
of environmental screening effects. Our results address a recent experimental observation of signatures of 
magnetic ordering in chiral graphene nanoribbons and provide an avenue towards tuning their 
properties via the structural and environmental degrees of freedom. 
\end{abstract}

\maketitle

%

Graphene and derived nanostructures exhibit a large number of novel electronic properties \cite{Geim07,CastroNeto09}. 
One of such features is the presence of electronic states localized at the edges of 
this two-dimensional (2D) nanomaterial \cite{Nakada96}. These zero-energy edge states were predicted 
to give rise to a novel type of magnetic ordering \cite{Fujita96} which may lead to practical 
carbon-based magnetic materials \cite{Makarova05}. Even more appealing is the prospective of 
realizing the theoretical proposals of novel spintronic devices based on graphene \cite{Son06,Yazyev08b,Wimmer08,Munoz-Rojas09,Yazyev10b}. 
While evidence for edge states has been seen experimentally \cite{Kobayashi05,Kobayashi06}, 
no solid proof of low-dimensional edge magnetism in graphene was presented until now.

%

A recent scanning tunneling microscopy/spectroscopy study of graphene nanoribbons (GNRs) with 
ultrasmooth edges showed the presence of edge states with characteristic splitting in the $dI/dV$ spectra -- an 
unambiguous indication of magnetic ordering \cite{Tao10}. These GNRs produced by the chemical unzipping
of carbon nanotubes \cite{Jiao10} are chiral, {\it i.e.} characterized by low-symmetry orientation of the edges
rather than by high-symmetry zigzag and armchair directions. While the presence of edge states at the chiral 
graphene edges is broadly recognized \cite{Nakada96,Akhmerov08,Wimmer10}, theoretical investigations of magnetic 
ordering driven by electron-electron ($e$--$e$) interactions have so far focused only on high-symmetry zigzag edges.     

%

In this Letter, we systematically study the electronic structure of chiral GNRs using 
a self-consistent model Hamiltonian approach including $e$--$e$ interactions. 
In particular, we investigate GNRs characterized by a broad range of chiralities and widths 
as well as address the effects of varying $e$--$e$ interaction strength.
Our study reveals that spin-polarized edge states are an intrinsic feature of chiral
GNRs, in agreement with the recent experimental observations. 
Moreover, we find a number of structure-property relations and unambiguous 
signatures of magnetic ordering of edge states which open new prospectives for their 
further exploration and for developing practical spintronic devices based on them.

%

\begin{figure}[b]
\includegraphics[width=7.8cm]{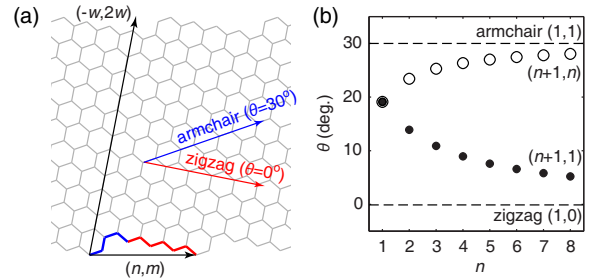}
\caption{\label{fig1}
(color online). (a) Atomic structure of a model of chiral graphene nanoribbon. The structure shown 
corresponds to $\theta = 10.9^\circ$ chiral GNR characterized by edge repeat vector
(4,1) and width $w=6$. Zigzag and armchair units of the edge are indicated. (b) Chirality angles 
$\theta$ of the considered ($n+1$,1) and ($n+1$,$n$) series of chiral GNRs.
}
\end{figure}

\begin{figure*}
\includegraphics[width=13.5cm]{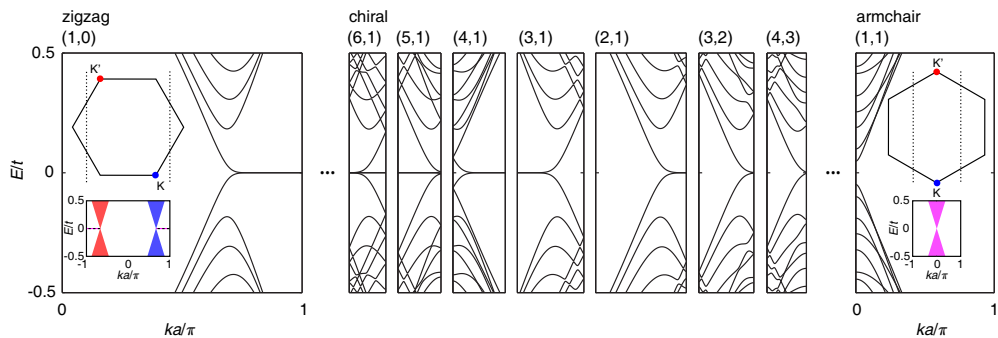}
\caption{\label{fig2}
(color online). Evolution of the tight-binding band structures (no $e$--$e$ interactions) of graphene nanoribbons ($w = 12$) upon 
the change of chirality from zigzag ($\theta = 0^\circ$) to armchair ($\theta = 30^\circ$) via the series 
of intermediate chiral configurations. The scales of the plots account for the varying Brillouin-zone 
dimensions. The insets show the relation between the one-dimensional band structures of zigzag and 
armchair GNRs and the two-dimensional band structure of ideal graphene.
}
\end{figure*}

The method employed in our study is based on the mean-field approximation to the
Hubbard Hamiltonian
\begin{align}
\label{eq1}
	{\mathcal H} =& - t \sum_{\langle i,j \rangle, \sigma} [ c_{i\sigma}^\dagger c_{j\sigma} + {\rm h.c.} ] + \nonumber \\ 
	              & + U \sum_i \left( n_{i\uparrow} \langle n_{i\downarrow} \rangle  + \langle n_{i\uparrow} \rangle n_{i\downarrow} - \langle n_{i\uparrow} \rangle \langle n_{i\downarrow} \rangle \right). 
\end{align}
The first term is the nearest-neighbor tight-binding Hamiltonian in which $c_{i\sigma}$ ($c_{i\sigma}^\dagger$) 
annihilates (creates) an electron with spin $\sigma$ at site $i$, $\langle i,j \rangle$ 
stands for the nearest neighbor pairs of atoms, and $t \sim 3$~eV \cite{CastroNeto09}. 
The second term accounts for $e$--$e$ interactions. The expectation values of the spin-resolved density 
$n_{i\sigma}= c_{i\sigma}^\dagger c_{i\sigma}$ depend on the eigenvectors of ${\mathcal H}$. Thus,
a self-consistent field procedure is used for solving the problem. The magnitude of the on-site 
Coulomb repulsion parameter $U/t \sim 1$ was estimated on the basis of first-principles calculations
and experimental data \cite{Yazyev08}. However, we point out that this effective parameter may also incorporate 
environmental factors such as the enhancement of screening due to the proximity of dielectric or metallic 
substrate \cite{Tao10}. For this reason we also study the dependence of results on $U/t$.

%

Two parameters determine the structure of smooth GNRs: (i) the crystallographic 
direction of the edge, and (ii) the width. In general, the direction of the nanoribbon's 
edge is defined by some translation vector ($n$,$m$) of the graphene lattice. For high-symmetry 
zigzag and armchair edges these vectors are (1,0) and (1,1), respectively. Provided that $n > m$,
the edge of ($n$,$m$) nanoribbon is a repeating structure composed of $n-m$ zigzag units and $m$ 
armchair units as illustrated for the particular case of a (4,1) GNR in Fig.~\ref{fig1}(a).
The length of repeat vector $a=\overline{(n,m)}=a_0\sqrt{n^2+nm+m^2}$, where $a_0 = 0.246$~nm is the 
lattice constant of graphene. Alternatively, chirality can be described by the chirality
angle $\theta=\arcsin \sqrt{ \frac{3}{4} \left ( \frac{m^2}{n^2+nm+m^2} \right )} $. 
Zigzag and armchair edges are characterized by $\theta = 0^\circ$
and $\theta = 30^\circ$, respectively, while for chiral edges $0^\circ < \theta < 30^\circ$.
We consider chiral GNRs defined by edge translational vectors ($n+1$,1) and 
($n$,$n+1$) ($n \ge 1$). These two series cover the whole range of chirality angles $\theta$ and
converge to $\theta = 0^\circ$ and $\theta = 30^\circ$, respectively, with increasing $n$ 
[Fig.~\ref{fig1}(b)]. The configurations of chiral GNRs considered in this study cover 
$4.7^\circ < \theta < 25.3^\circ$. The width $W = \sqrt{3}w a_0 \cos \theta$ of a GNR 
is defined by vector ($-w$,$2w$) pointing along the armchair direction as shown in Fig.~\ref{fig1}(a).


We start our discussion by considering the evolution of tight-binding band structures of GNRs (neglecting $e$--$e$ interactions, $U/t=0$)
upon the change of chirality from zigzag ($\theta = 0^\circ$) to armchair ($\theta = 30^\circ$) via the 
series of intermediate chiral configurations at fixed nanoribbon width ($w = 12$). High-symmetry 
zigzag nanoribbon [Fig.~\ref{fig2}, left panel] exhibits a flat band at the Fermi level ($E=0$) which spans one-third
of the 1D Brillouin zone (BZ), that is, the corresponding density of edge states per edge length 
per spin $\rho(\theta = 0^\circ) = 1/(3a_0)$. Armchair GNRs [Fig.~\ref{fig2}, right panel] are 
either metals or semiconductors with no electronic states localized at the edges. This result can 
be rationalized within the infinite-width picture by considering the 2D band structure of graphene 
projected onto the 1D BZ of periodic edge. In the case of an armchair GNR both Dirac points ($K$ and $K'$)
of graphene's band structure are projected onto the $\Gamma$ point of 1D BZ.
However, for a zigzag GNR points $K$ and $K'$ are projected onto $k = 2\pi/(3a)$ and $k = -2\pi/(3a)$, respectively, 
with zero-energy flat band connecting these two points [Fig.~\ref{fig2}, insets]. The band structures 
of chiral GNRs in the infinite-width limit can be obtained by continuous rotation of the band structure
of graphene which leads to the known result \cite{Akhmerov08}:
\begin{equation}
\label{eq2}
 \rho(\theta) = \frac{2}{3a_0} \cos \left( \theta + \frac{\pi}{3} \right).
\end{equation}
That is, in the limit of large width only armchair GNRs show no edge states. The density of edge states is
largest for zigzag nanoribbons and shows almost linear dependence on $\theta$. The tight-binding band structures
of $w = 12$ chiral GNRs [Fig.~\ref{fig2}; the $x$-axis scales correspond to the 1D BZ dimensions]
confirm this picture. Two important comments should be made: (i) in finite-width GNRs with edge
orientation close to the armchair direction zero-energy edge states are partially or even completely
suppressed; (ii) in GNRs with edge orientation close to the zigzag direction the flat edge-state band spans
whole 1D BZ and becomes multiple degenerate due to band folding. 


\begin{figure}
\includegraphics[width=7.8cm]{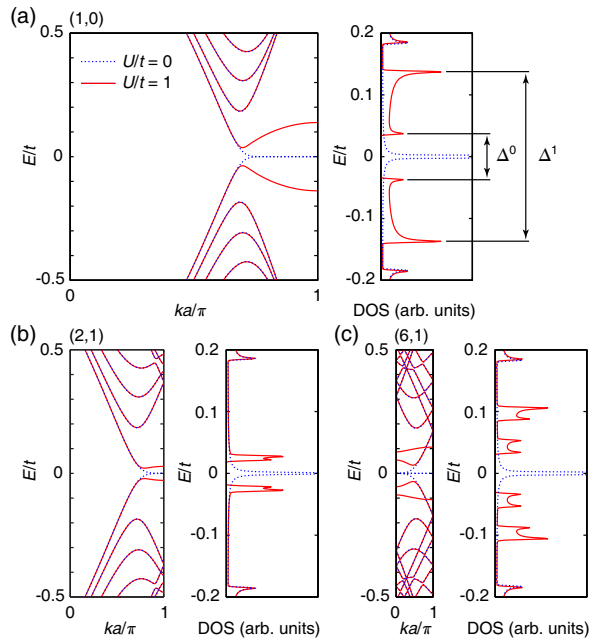}
\caption{\label{fig3}
(color online). (a) Effects of the electron-electron interactions on the band structure (left panel)
and the density of states (right panel) of a zigzag GNR ($w = 12$). Dashed and solid
curves correspond to the tight-binding model ($U/t=0$) and the mean-field Hubbard model ($U/t=1$), respectively.
(b),(c) Respective plots for (2,1) and (6,1) chiral GNRs. 
}
\end{figure}

We now discuss the effects of $e$--$e$ interactions on the electronic spectra of chiral GNRs.
Figure~\ref{fig3}(a) shows the band structure and the density of states (DOS) plot for a zigzag GNR 
($w = 12$) obtained within the mean-field Hubbard model ($U/t=1$) compared to the tight-binding
results ($U/t=0$). In the presence of $e$--$e$ interactions the electronic ground state of the zigzag
GNR exhibits an interesting magnetic ordering: ferromagnetic (FM) along the edges and antiferromagnetic (AFM) 
across the nanoribbon \cite{Fujita96}. The magnetic moment per edge unit length $M = 0.313 \mu_{\rm B}/a_0$. 
Spin-polarization lifts the degeneracy of edge states and opens an electronic band gap $\Delta^0$. While the tight-binding DOS has only one van Hove 
singularity related to the presence of 1D edge states at $E = 0$, the mean-field Hubbard model solution
shows two pairs of peaks split by $\Delta^0$ and $\Delta^1$. We note that Hamiltonian (\ref{eq1}) respects electron-hole symmetry.
Splitting $\Delta^0$ is related to magnetic correlation between the opposite edges while the larger splitting $\Delta^1$ 
is due to the FM correlation along one single edge of most
strongly localized electronic states at $k = \pi/a$ \cite{Son06,Fernandez-Rossier08,Jung09}. Thus, the
two splittings $\Delta^0$ and $\Delta^1$ constitute independent signatures of magnetic ordering across 
the nanoribbon and along its edge. The rest of electronic spectrum ($|E| > 0.18t$) is negligibly affected by the Hubbard
term. Van Hove singularities at $|E| \approx 0.2t$ and higher energies correspond to the bulk-like
states subjected to quantum confinement in 1D GNRs of the given width. 
The mean-field Hubbard model electronic spectra of chiral GNRs show all the features characteristic of
zigzag GNRs provided the ground state is spin-polarized. Both splittings, especially $\Delta^1$, are 
reduced in the case of (2,1) GNR ($\theta = 19.1^\circ$) due to the smaller magnetic moment
$M = 0.096 \mu_{\rm B}/a_0$ [Fig.~\ref{fig3}(b)]. Nanoribbons with chirality close to $\theta = 0^\circ$
show additional pairs of van Hove singularities due to splittings of the multiple degenerate
edge-state bands at $k=0$ and $k=\pi/a$ [Fig.~\ref{fig3}(c) for $\theta = 7.6^\circ$ (6,1) GNR].
 

\begin{figure}[b]
\includegraphics[width=7.8cm]{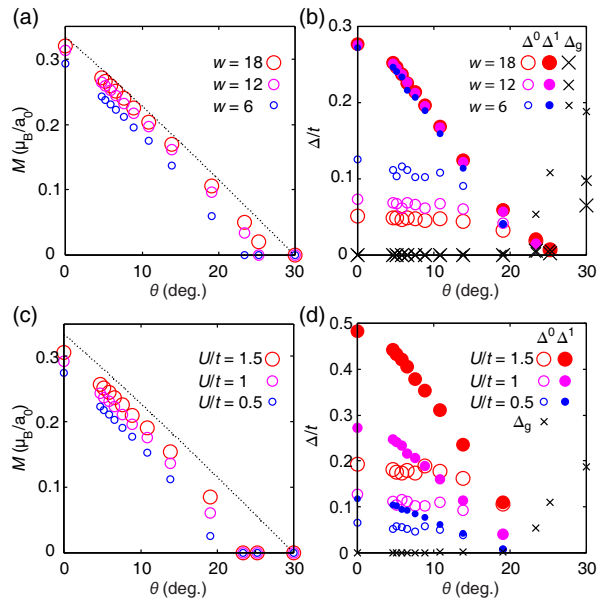}
\caption{\label{fig4}
(color online). (a) Magnetic moment per edge unit length $M$ as a function of chirality angle $\theta$
for three different nanoribbon widths $w$ from the mean-field Hubbard model ($U/t=1$) calculations. 
The dotted line shows magnetic moments in the limit of infinite width [Eq.~(\ref{eq2})]. (b) Electronic band gap 
$\Delta^0$ and maximum energy splitting $\Delta^1$ as a function of $\theta$ for different values of 
$w$ (mean-field Hubbard model, $U/t=1$). Crosses indicate the tight-binding band gaps $\Delta_{\rm g}$. 
The values of (c) $M$, (d) $\Delta^0$ and $\Delta^1$ obtained using the mean-field Hubbard model
at different values of $U/t$ and $w=6$.
}
\end{figure}

Figure~\ref{fig4} summarizes the calculated magnetic moments and energy splittings for GNRs of various
chiralities and widths, and different magnitudes of $U/t$. At $U/t = 1$ spin-polarized ground states of 
few-nm-wide GNRs span almost full range of chiralities [Fig.~\ref{fig4}(a)]. The magnetic moments per unit 
length $M$ follow closely the dotted curve which corresponds to the complete spin-polarization of edge states in the 
infinite-width limit (\ref{eq2}), but appear to be negatively shifted by nearly constant amounts which
are moderately dependent on width. Splitting $\Delta^1$ shows a similar dependence on $\theta$ and it is independent
of $w$ [Fig.~\ref{fig4}(b)]. On the contrary, the magnitude of $\Delta^0$ is largely 
insensitive to the variations of $\theta$ in broad ranges of this parameter, but shows a clear dependence on $w$
analogous to the case of zigzag GNRs \cite{Son06b}. 
As the chiral angle $\theta$ approaches $30^\circ$ magnetic moments vanish and the tight-binding band gaps 
$\Delta_{\rm g}$ quickly rise reaching their maximum values for the corresponding armchair GNRs \cite{Ezawa06}.
Figure~\ref{fig4}(c) reveal a moderate dependence of $M(\theta)$ on the strength of $e$--$e$ 
interactions. However, the magnitudes of $\Delta^0$ and $\Delta^1$ are both approximately proportional to $U/t$ 
[Fig.~\ref{fig4}(d)].  

%

\begin{figure}
\includegraphics[width=7.8cm]{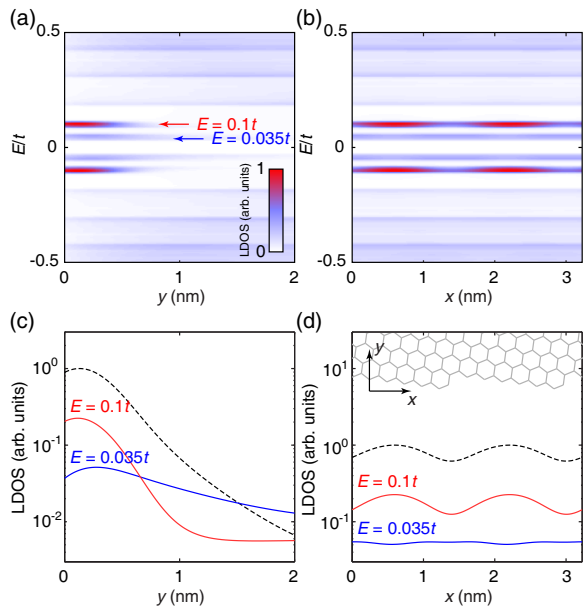}
\caption{\label{fig5}
(color online). Variation of the local density of states (LDOS) (a) across the (6,1) chiral GNR
($x$-axis is oriented along the edge; $y = 0$ corresponds to the outermost edge atom) and (b) along its edge obtained using the mean-field 
Hubbard model ($U/t=1$). (c),(d) Log-linear plots of LDOS at the energies indicated in panel (a) across the 
(6,1) chiral GNR and along its edge, respectively. The dashed lines correspond to the tight-binding 
LDOS at $E=0$. The inset in panel (d) superimposes the edge structure with the plot.
}
\end{figure}

Finally, we turn our attention to the spatial variation of electronic spectra of magnetic graphene edges
in relation to the experimental observations \cite{Tao10}. As a case study we investigate the 
mean-field Hubbard model ($U/t = 1$) local density of states (LDOS) evaluated across the (6,1) chiral
GNR ($w = 12$, $W = 5$~nm) [Fig.~\ref{fig5}(a)] and along its edge [Fig.~\ref{fig5}(b)]. We find that 
both pairs of contributions to the total DOS due to the edge states seen in  Fig.~\ref{fig3}(c) are localized 
at the edge. However, the higher-energy LDOS peak at $E = 0.1t$ decays very fast being confined within the 
1-nm-wide edge region while the lower-energy feature at $E = 0.035t$ penetrates deep into the middle of GNR.
The series of peaks at $E \gtrsim 0.2t$ repeated in energy by $\approx 0.1t$ correspond to the bulk-like graphene states
subjected to quantum confinement. The total density of edge states from the tight-binding 
calculations shows pronounced oscillations along the GNR edge [dashed curve in Fig.~\ref{fig5}(d)].
The oscillation period corresponds to the edge periodicity $a = 1.6$~nm. The $E = 0.1t$ LDOS peak 
from the mean-field Hubbard model follows this trend. In contrast, the lower-energy peak ($E = 0.035t$) shows
weak variations of LDOS. This behavior was found to be generic to all studied chiral and zigzag magnetic GNRs.


To summarize, our model Hamiltonian study of the electronic structure and magnetic properties 
of chiral graphene nanoribbons revealed a number of structure-property relations. The described 
relations can serve as unambiguous signatures of edge-state magnetism in graphene nanoribbons 
and provide an avenue towards controlling their magnetic and electronic properties. 


We would like to thank C.~Tao and M.~F.~Crommie for discussions. This work 
was supported by NSF Grant No. DMR10-1006184 (numerical simulations of GNRs) and by 
the Director, Office of Science, Office of Basic Energy Sciences, Division
of Materials Sciences and Engineering Division, US Department of Energy under
Contract No. DE-AC02-05CH11231 (software development of electron correlation effects). 
R.B.C. acknowledges financial support from Brazilian agencies CNPq, CAPES, FAPERJ and 
INCT – Nanomateriais de Carbono and the ONR MURI program (computer support).

\end{document}